\documentclass{article}
\usepackage{graphicx} 
\usepackage{amsmath}
\usepackage{times}
\usepackage{xcolor}
\usepackage{amssymb}
\usepackage{amsmath}
\usepackage{blindtext}
\usepackage{hyperref}
\usepackage{comment}

\title{Assessing the pair interactions of pNIPAM microgel particles using optical tweezers}
\author{Jos\'e Mu\~{n}et\'on-D\'iaz$^{1}$, Chi Zhang $^{1}$, Priti Sundar Mohanty $^{1,2}$, \\
Frank Scheffold $^{1  \star}$\\
\\
\small{$^{1}$Department of Physics, University of Fribourg, 1700 Fribourg, Switzerland}\\
\small{$^{2}$School of Biotechnology, KIIT Deemed to be University, Bhubaneswar, Odisha 751024, India}
\\
\small{$^\star$ Correspondence to frank.scheffold@unifr.ch}
}

\date{}
\begin{document}
\maketitle

\begin{abstract}
We experimentally study the pairwise interactions between poly(N-isopropylacrylamide) (pNIPAM) microgel particles using line optical tweezers. To measure their interaction potentials under controlled conditions, we trap two microgel particles in the tweezer and analyze their thermal positional fluctuations near contact. The pair interaction potential is modeled using a Hertzian core–polymer brush corona framework, capturing the complexity of particle interactions by accounting for both the deformable, elastic core and the steric effects of the polymer brush-like shell. Our experimental results demonstrate that the brush corona interactions soften as the system approaches the lower critical solution temperature (LCST) of the microgels, consistent with the decay of the polymer's second virial coefficient.
\end{abstract}

\section{Introduction}
Microgel particles are soft, polymer-based colloids that can swell or shrink in response to external stimuli such as temperature \cite{Pelton1986}, pH \cite{karg2008temperature}, or ionic strength \cite{mohanty2008structural}. Typically ranging from the micron to sub-micron scale, these particles consist of cross-linked polymer networks that absorb water, creating a gel-like structure. Their ability to respond to environmental changes classifies microgels as "smart" materials, which makes them highly versatile in applications such as drug delivery, tissue engineering, and soft robotics \cite{Guan2011}.
\newline The microscopic structure of microgel particles significantly influences their physical behavior and interaction potential~\cite{bergman2024understanding}. The degree of crosslinking controls particle stiffness and deformability. Additionally, the cross-linked polymer network often forms a core-shell structure, where the outer shell is much softer compared to the core, due to the crosslinker being consumed faster than the monomer during microgel synthesis. Recent advances have provided a more detailed understanding of this microscopic structure, offering insights into how microgels behave under various conditions \cite{scheffold2020pathways, ninarello2019modeling}. The soft, deformable nature of microgels, along with their brush-like outer layer, creates a complex interaction profile distinct from traditional hard colloids, allowing for compression, overlap, or even interpenetration in dense suspensions \cite{conley2017jamming}.
\newline The knowledge of the microscopic structural properties may help provide a comprehensive understanding of the properties of dense microgel assemblies. In particular, understanding their pair potential is crucial for interpreting how structural and mechanical properties manifest in their collective behavior. These interactions influence phase transitions, self-assembly, and the mechanical properties of microgel suspensions, particularly in crowded environments where particles experience non-trivial forces. Previous research has revealed the relationship between the microscopic structure of microgel particles, obtained through microscopy or modelling and their rheological properties when densely packed \cite{scheffold2010brushlike, bonnecaze2010micromechanics, conley2019relationship, ghosh2019linear}. In this work, we present direct dynamic measurements of the pair interactions of pNIPAM microgel particles using line optical tweezers (LOT). We model the weak repulsive interactions using a Hertzian core–polymer brush shell framework. This model effectively captures the elastic repulsion between weakly deformed microgels and provides insight into how microgels interact in dense suspensions or under external stimuli.

\section{Experiments}
We synthesized standard poly(N-isopropylacrylamide) (pNIPAM) microgel particles as previously described in ref.~\cite{mohanty2014effective}, using a N,N$^\prime$-Methylenebis(acrylamide) (BIS) crosslinker density of 5 mol\%. The size and core-shell structural parameters were determined from static light scattering (SLS) data (LS spectrometer V2, LS Instruments, Switzerland), where we employed the fuzzy sphere model with a particle radius $R \simeq R_\textrm{c} + 2\sigma$~\cite{stieger2004thermoresponsive}, as detailed in \cite{reufer2009temperature}. The results of this analysis are presented in Table\ref{tab:SLS}. From the fit, we find a size polydispersity of about 6-8\% (standard deviation divided by the mean).
\newline The pairwise interactions were measured using line optical tweezers (LOT) formed by a pair of cylindrical lenses ($f_{1} = 15$~mm and $f_{2} = 400$~mm). Briefly, a Gaussian laser beam is first reduced to a diameter of around $0.7$~mm using a 10X beam expander, then sent through the cylindrical lens telescope (in a $4f$ configuration \cite{biancaniello2006line}). The beam is expanded in one direction by the cylindrical lens pair and directed to the back aperture of the objective (Nikon Apo TIRF 100x 1.49 oil). This setup creates a line trap at the focal plane of the objective. The sample temperature is controlled using a stage-top heating/cooling chamber (Bold Line, Okolab, Italy).

\begin{figure}
	\centering
    \includegraphics[width=1\linewidth]{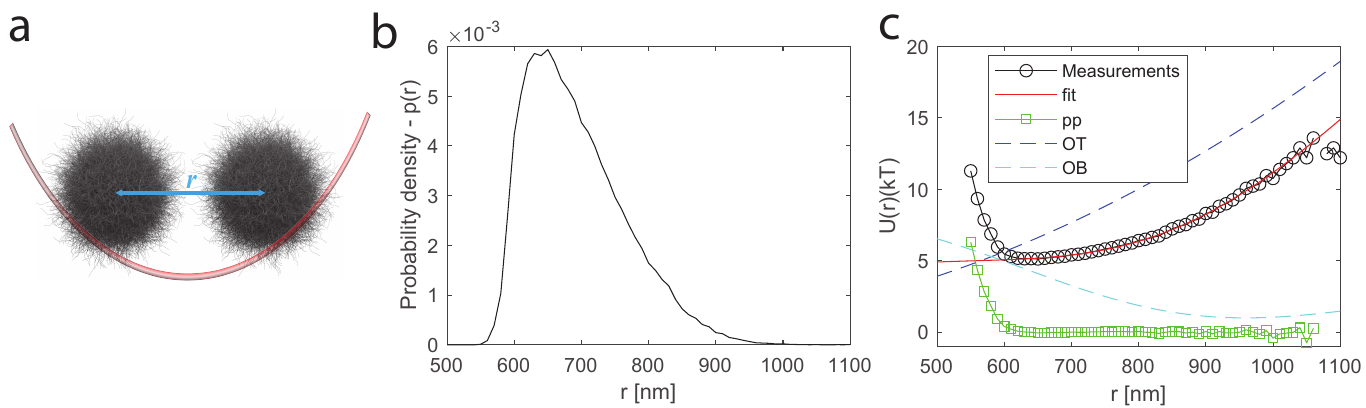}
	\caption{Sketch of two microgel particles in an optical line trap (parabolic potential). The center-to-center distance $r$ is determined via particle tracking. The center-to-center distance histogram $P(r)$, shown in panel (b), is influenced in a characteristic way by both the optical potential applied to the particles and their intrinsic pairwise interactions. In the measured potential (black circles in panel c), the contributions from the trap (blue dashed line) and the optical binding (light blue dashed line) can be removed following the method described in ref.~\cite{zhang2024determining}. The fitted optical potential $U_{OT} + U_{OB}$, shown as a red solid line in panel c), is then subtracted, revealing the particle-particle intrinsic interaction (green squares).}
	\label{fig:LOT}
\end{figure}

\begin{table*}[h]\centering
\resizebox{\textwidth}{!}{
\begin{tabular}{r|rrrrrrrrrr}
Temperature ($^\circ$C) 	& 20 & 22 & 24 & 26 & 28 & 30 & 32 & 34 & 36 & 38\\
\hline
$R$ (nm) & 524.4&509.5	&495.0&	476.5&	450.2	&415.4&	365.6	&313	&309 & 304\\
$R_\textrm{c}$  (nm)  &  401.8 & 393.3 & 385.4 &  375.1 & 361.4 & 342.8 & 320.1 & 313 & 309 & 304\\
$\sigma$ (nm)& 61.3 & 58.1 & 54.8 & 50.7 & 44.4 & 36.3 & 22.7 & -- & -- & --\\
\end{tabular}}\caption{Static light scattering (SLS) analysis: sizes of microgel particles measured by SLS at different temperatures, fitted to the fuzzy-sphere model. The measured sizes suggest a lower critical solvent temperature $T_\text{LCST}$ around $ 33^\circ \mathrm{C}$.}
\label{tab:SLS}
\end{table*}
Data acquisition, particle tracking, and pair potential calculations were carried out following the method described in detail in our previous study \cite{zhang2024determining}. Briefly, two microgel particles were brought in close proximity using the LOT, as illustrated in the sketch shown in Figure\ref{fig:LOT} (a). The images were recorded with an sCMOS camera (Prime 95B, Teledyne Photometrics, US). The residual errors in the position measurements can be assessed by looking at the relative motion of the particle pair, as shown in our previous study~\cite{zhang2024determining}. 
We selected an exposure time of 25 $\mathrm{\mu s}$ as a good compromise regarding the tradeoff between dynamic error, due to motion artifacts, and static error dictated by signal-to-noise. The dynamic error for measurements performed at a temperature of 22$^\circ$C is calculated to be around 4–5 nm. The static error is measured to be around 5 nm, such that the total error is $\sim \sqrt{50} \, \mathrm{nm} \approx 7 \, \mathrm{nm}$. 
\newline With the help of particle tracking using full image reconstruction, as explained in \cite{zhang2024determining}, the center-to-center distance of the particle pair is extracted. A typical position histogram is shown in Figure\ref{fig:LOT} (b). Subsequently, the histogram is converted to a measured potential $U(r)$ using the Boltzmann distribution for the probability density distribution $P(r) \propto \exp[-U(r)/k_B T]$ (see Figure\ref{fig:LOT} (c)), which contains the intrinsic pair potential (pp), the parabolic potential of the trap (OT), and the optical binding potential (OB) arising from the complex scattering between the particles. A comprehensive fitting procedure following~\cite{zhang2024determining} allows us to remove the contributions to the potential due to otpical forces and gives access to the intrinsic particle-particle interaction.
\section{Modelling and fitting}
We model the microgel particles as having a deformable solid core and a polymer brush shell \cite{scheffold2010brushlike}. The soft core is described as a Hertzian elastic sphere, with its interaction potential given by
\begin{equation} \label{eq:Hertzian} U_\text{c}(\Delta_{c}) = \frac{2Y(2R_{c})^{3}}{15(1-\nu^{2})} \left( \frac{\Delta_{c}}{2R_{c}} \right)^{2.5}, \end{equation}
where $Y$ is Young's modulus, $\nu$ is Poisson's ratio, and $\Delta_{c}$ is the core deformation. For a slightly compressed sphere, which is the case in the optical tweezers measurement, the single Hertzian model accurately describes the relationship between the energy and deformation. However, when the microgel particles are more significantly compressed, one would expect that Young's modulus increases toward the center of the particle, as discussed in ref.~\cite{bergman2018new}, where the authors modeled particle interactions using a multi-layer Hertzian model.
\newline We model the outer shell as a polymer brush. The approach to modeling repulsive interactions between brush-coated surfaces is based on the scaling theory for polymer brushes, initially developed by Alexander and de Gennes~\cite{alexander1977adsorption, de1980conformations}. This model captures the repulsive forces between two surfaces coated with polymer brushes and has been applied to understand the elasticity of microgel coronas in good solvent conditions \cite{scheffold2010brushlike}. In this work, we employ the mean-field polymer theory summarized by Spencer and coworkers in ref.~\cite{espinosa2013impact} to address brush swelling and elasticity. Using this approximation, the polymer brush is characterized by a parabolic density profile. By applying the Derjaguin approximation, the potential between two brush-coated spherical surfaces can be expressed as follows:
\begin{equation}
\label{eq:EspinosaEnergyChi}
    \begin{split} U_\text{s}(\Delta_{s})=
    C\left(T\right)\frac{R_{c}}{30 }  \left(u^6-10 u^3+54 u-30 \log (u)-45\right),
\end{split}
\end{equation}
where $u=\Delta_{s}/L\le1$, with $\Delta_{s}$ being the deformation of the shell with a equilibrium thickness $L$. In the equation \ref{eq:EspinosaEnergyChi}, 
\begin{equation}
\label{eq:CT}
    C\left(T\right)=4\pi N_{p}^2 a^3/s^4 \times \tau\left(T\right)
\end{equation}
quantifies the energy penalty of compressing the brush. The effect of temperature is captured by the dimensionless virial coefficient $\tau \left(T\right)= T_\text{LCST}/T - 1 \simeq 1 - T/T_\text{LCST}$, where $\tau \left(T\right) \propto A_2\left(T\right)$ is proportional to the experimentally accessible second virial coefficient, as reported by Kubota et al. in ref.~\cite{kubota1990single}. As the swollen microgel system approaches the lower critical solution temperature (LCST), the virial coefficient decreases, leading to a softening of the polymer brush.
\newline We model the Hertzian core coated with a polymer brush as two coupled elastic springs in series. When a repulsive force is established due to microgel interaction (for $u < 1$), the compressional force on both the core and shell is the same, but they deform by different amounts. The deformations of the core \(\Delta_{c}\) and shell \(\Delta_{s}\) depend on their respective properties, as described by equations~\eqref{eq:Hertzian} and \eqref{eq:EspinosaEnergyChi}. Given a total deformation \(\Delta=2R-r\), \(\Delta_{c}\) and \(\Delta_{s}\) are determined by equating the forces derived from the derivatives of $U_\text{c}(\Delta_{c})$ and $U_\text{s}(\Delta_{s})$. This allows us to compute the potential \(U(r)\), which can then be compared with the measured pair potential from optical tweezers, enabling parameter fitting for equations \eqref{eq:Hertzian} and \eqref{eq:EspinosaEnergyChi}.
\newline To achieve a stable fit of the measured pair potential, it is necessary to fix certain parameters. We use information from static light scattering (SLS) to fix the characteristic ratio of $R_\text{c}$ and $R=R_\text{c}+L$, sometimes denoted as $\alpha$ where $L=2\sigma$, Table \ref{tab:SLS}.  The optical tweezer measurements are highly sensitive to the global particle size, $R$, as it determines the onset of interactions. Since the individual particles studied are drawn from a population with finite polydispersity, we treat the absolute value of $R$ as a fitting parameter.
\begin{figure}[ht]
	\centering
    \includegraphics[width=0.8\linewidth]{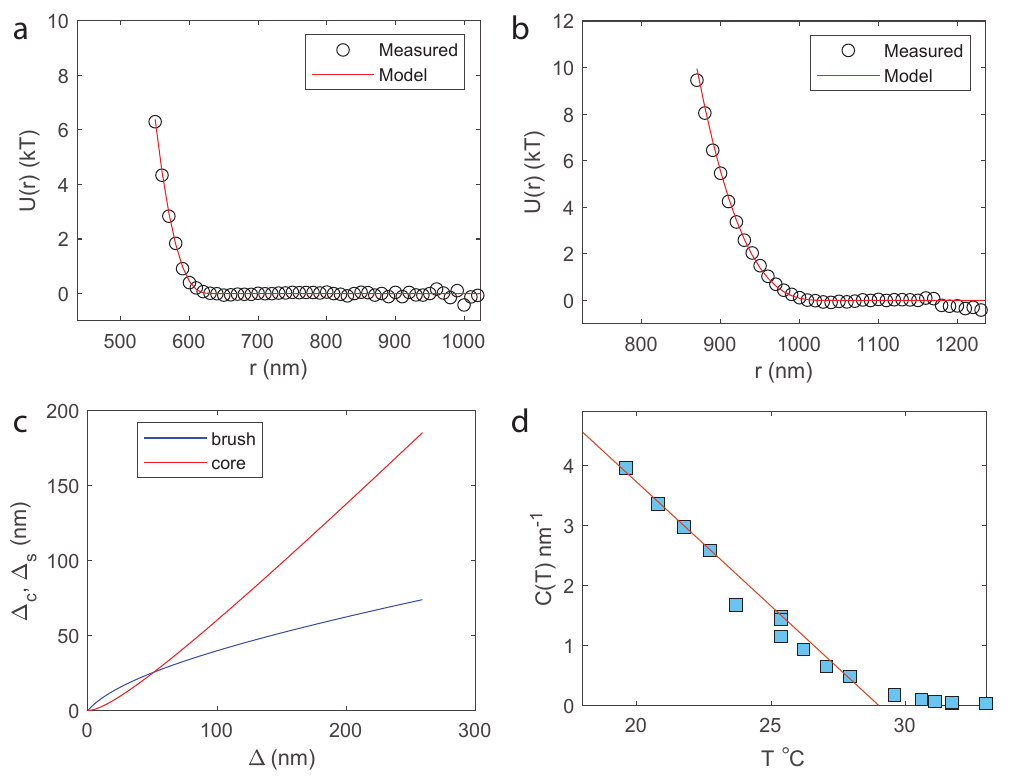}
	\caption{Pair potential measured using line optical tweezers at $T=36^\circ \mathrm{C}$ (a) and $T=22^\circ \mathrm{C}$ (b). Black circles represent the measured values. Red lines show the fit to the model, which is a pure Hertzian type in (a) and a Brush-Hertzian coupled type in (b). The deformation of the core and shell as a function of the total deformation at $T=22^\circ \mathrm{C}$ is presented in (c). Panel (d) shows the temperature dependence of the brush elasticity parameter extracted from the fit. The red solid line indicates a linear decay, fit with $C\left(T\right)=125/\mathrm{nm} \times \left(1- T/\left(302.15\mathrm{K}\right)\right)$.}
	\label{fig:Potentials}
\end{figure}
Young’s modulus $Y$ of the core changes as the particles swell when lowering the temperature. As discussed in ref.\cite{yasuda2020universal}, we can estimate the core modulus as follows. At a reference temperature $T_{0}$, above $T_\text{LCST}$, where the microgel particles are fully collapsed, the pair potential arises solely from the core interaction since the brush-like shell is collapsed onto the core. Thus, we can determine the Young’s modulus in this collapsed state ($Y_0$) by fitting the measured potential to a pure Hertzian model (equation \ref{eq:Hertzian}). As the temperature decreases and the core expands, the corresponding Young’s modulus can be estimated as $Y(T) = Y_0 (R_\text{c}(T_{0}) / R_\text{c}(T))^3$. With these parameters set, the remaining fitting parameters are the total particle radius $R$ and the brush elasticity parameter $C(T)$.
\section{Results} We first analyze the pair potential measured at $36^\circ \mathrm{C}$, shown in Figure\ref{fig:Potentials}~(a). Since the brush shell is fully collapsed, the microgel particles can be treated as homogeneous soft spheres, and we model the interaction potential using equation~\eqref{eq:Hertzian}, setting $U_s \equiv 0$. We find that Young's modulus at $T=36^\circ \mathrm{C}$ is $128$ Pa, and the measured particle radius is $R = 312$ nm in good agreement with the values reported in Table~\ref{tab:SLS}.
Next, Figure \ref{fig:Potentials} (b) shows the measured pair potential  at $22 ^\circ \mathrm{C}$. Both the particle radius $R$ and the brush elasticity $C(T)$ can be extracted from the fit. The data shows good agreement with the experimental values, yielding $R = 518$ nm, $C(T=22 ^\circ ) = 2.97$nm$^{-1}$ with  $Y(T=22 ^\circ ) = 54.7$~Pa. In panel (c), for the same temperature, we show the deformation of the core and shell as a function of total compression $\Delta$. Initially, during weak compression, the brush layer contributes more significantly, but as compression increases, the core becomes dominant. In the tweezer experiments, we only probe deformations $\Delta \leq 100$ nm, as thermal energies are insufficient to explore stronger deformations. Actively pushing the particles together is not feasible because the line trap is not strong enough to keep the particles on-axis.
In Figure \ref{fig:Potentials} (d), we display the brush elasticity parameter $C(T)$ as a function of temperature. Consistent with equation \ref{eq:CT}, the brush softens as it approaches the lower critical solution temperature (LCST), $T_\text{LCST}$. Our experiments capture this behavior well, with the fitted brush elasticity decaying according to $C(T) = 125/\mathrm{nm} \times (1 - T / 302.15 , \mathrm{K})$, suggesting a $T_\text{LCST} = 29^\circ \mathrm{C}$.
\section{Summary and Conclusions}
We have demonstrated that the pair interaction of pNIPAM microgel particles can be extracted using optical tweezers. A simple model based on a Hertzian elastic core and a polymer brush shell describes our data well. No additional repulsive interaction contributions needed to be considered in the weak deformation regime—neither core stiffening nor electrostatic double-layer repulsion was observed. The situation might differ at lower ionic strength or for microgels with a higher surface charge density. Stronger compression would certainly lead to higher core elasticity. Our experimental findings align well with polymer theory, showing that the brush softens as the system approaches $T_\text{LCST}$. The experiments revealed a slight deviation in the LCST measured by optical tweezers (29 $^\circ$C) compared to that suggested by static light scattering (33 $^\circ$C), an observation that warrants further investigation. Our work highlights the power of quantitative optical tweezer measurements for capturing the interaction potential of microgels and provides valuable insights into their elasticity across different temperature regimes.

\bibliographystyle{unsrt}

\section*{Appendix}
Measurements of interaction potentials and fitted curves, with fitted parameters.\\

\begin{figure}
	\centering
    \includegraphics[width=1\linewidth]{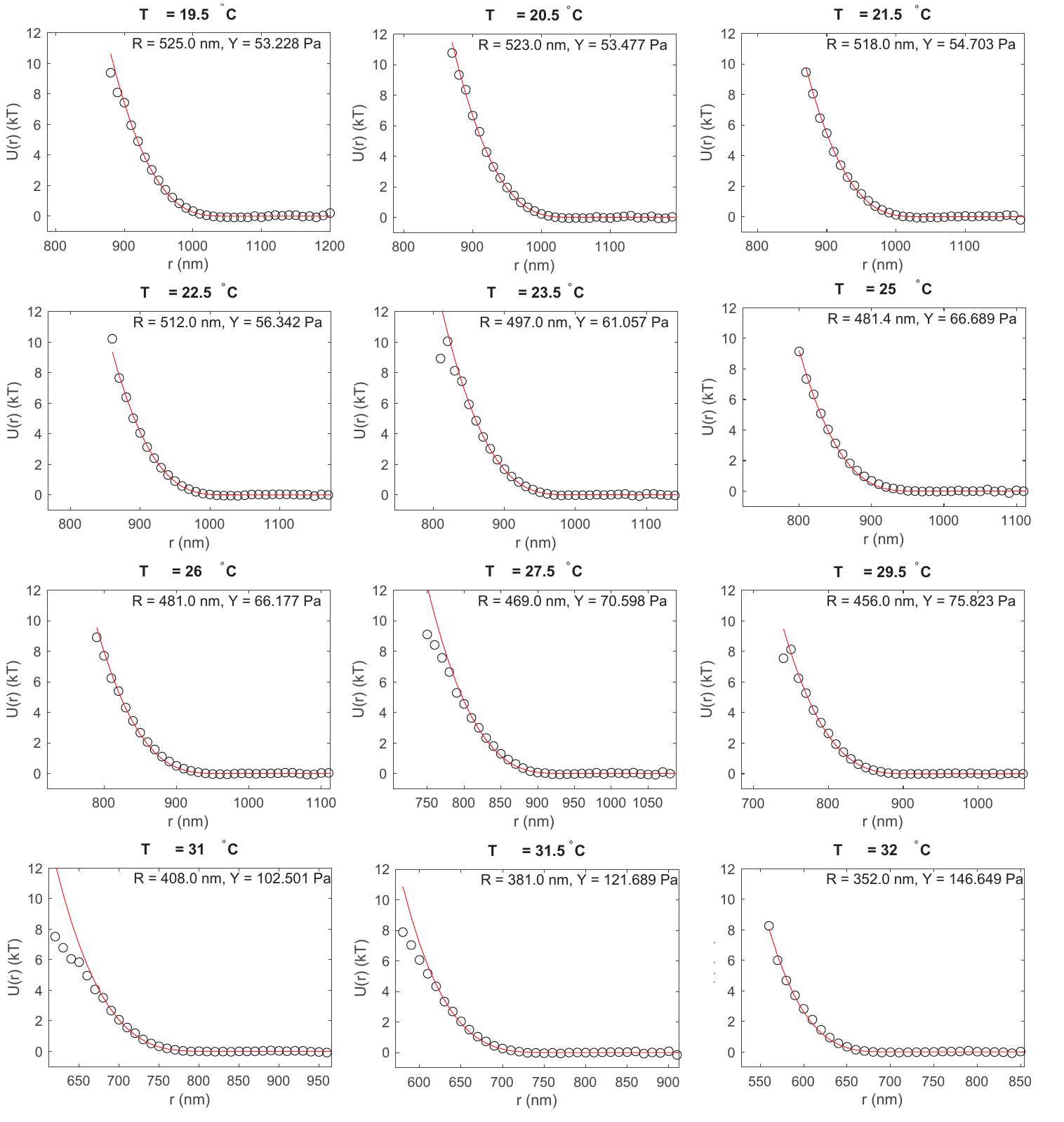}
	\caption{Measurements of pair potential reported in Figure\ref{fig:Potentials} using line optical tweezers. Black circles presents the measured value. Red lines show the fit to the Brush-Hertzian model.}
	\label{fig:AllPotentials}
\end{figure}
\end{document}